\documentclass[12pt]{article}

\usepackage{amssymb}
\usepackage{amsthm}

\usepackage{logique}

\usepackage{graphicx}

\newcommand{\NN}{\mathbb{N}}

\newcommand{\RR}{\mathbb{R}}

\newcommand{\AAA}{\mbox{\sffamily A}}
\newcommand{\BBB}{\mbox{\sffamily B}}
\newcommand{\CCC}{\mbox{\sffamily C}}

\newcommand{\III}{\mbox{\sffamily I}}

\newcommand{\KKK}{\mbox{\sffamily K}}

\newcommand{\WWW}{\mbox{\sffamily W}}

\newcommand{\pp}{\mbox{\sffamily p}}
\newcommand{\qq}{\mbox{\sffamily q}}

\newcommand{\gB}{{\cal B}\hspace{-0.65em}{\cal B}}

\newcommand{\indi}{^{\mbox{\footnotesize int}}}

\newcommand{\forces}{\raisebox{-0.2ex}{ $\Vdash$ }}
\newcommand{\nforces}{\raisebox{-0.2ex}{ $\nVdash$ }}

\newcommand{\PL}{\mbox{\sffamily PL}}

\newcommand{\ppt}{{\scriptstyle <}}
\newcommand{\lbr}{\langle}
\newcommand{\rbr}{\rangle}
\newcommand{\gl}{\gimel}

\newcommand{\ZFe}{ZF$_\varepsilon$}

\newtheorem{theorem}{Theorem}
\newtheorem{lemma}[theorem]{Lemma}
\newtheorem{corollary}[theorem]{Corollary}
\newtheorem{proposition}[theorem]{Proposition}

\author{Jean-Louis Krivine\\
}

\title{Bar recursion in classical realisability~:\\
dependent choice and continuum hypothesis}
\date{\footnotesize {\today}}

\begin{document}
\maketitle\noindent

\section*{Introduction}\noindent
This paper is about the bar recursion operator \cite{Spc}, in the context of classical realizability \cite{JLK2,JLK3}. It is a sequel to
the three papers~\cite{BBC,BO,Str}. We use the definitions and notations of the theory of classical realizability
as expounded in~\cite{JLK1,JLK2,JLK3}.\\
In \cite{BBC}, S.~Berardi, M. Bezem and T. Coquand have shown that a form of the bar recursion operator can be used,
in a proof-program correspondence, to interpret {\em the axiom of dependent choice} in proofs of $\Pi^0_2$-formulas
of arithmetic. Their work was adapted to the theory of domains by U. Berger and P. Oliva in \cite{BO}.
In \cite{Str}, T. Streicher has shown, by using the bar recursion operator of \cite{BO}, that the models of ZF\/, 
associated with realizability algebras~\cite{JLK1,JLK3} obtained from usual models of $\lbd$-calculus
(Scott domains, coherent spaces,~\ldots), satisfy the axiom of dependent choice.\\
We give here a proof of this result, but for a realizability algebra which is built following the presentation
of~\cite{BBC}, which we call the BBC-algebra.\\
In section~\ref{BBCalg}, we define and study this algebra~; we define also the bar recusion operator,
which is a closed $\lbd$-term.\\
In sections~\ref{chden} and~\ref{chdep}, which are very similar, we show that this operator 
realizes the axiom of countable choice (CC), then the axiom of dependent choix (DC). The proof is
a little simpler for CC.\\
In section~\ref{WOR}, we deduce from this result, using results of \cite{JLK4} that, in the model of ZF associated
with this realizability algebra,
\emph{every real (more generally, every sequence of ordinals) is constructible}.\\
The formulas ``$\mathbb{R}$\emph{ is well ordered}''\ and ``\emph{Continuum hypothesis}'' are therefore
realized, in these models, by closed $\lbd_c$-terms (i.e. $\lbd$-terms containing the control instruction
$\ccc$ of Felleisen-Griffin).\\
We show also that \emph{every true formula of analysis is realized by a closed $\lbd_c$-term}.\\
In this way, we show how to obtain a program (closed $\lbd_c$-term) from any proof of a $\Pi^0_2$ arithmetical
formula  in the theory \ ZF + ``Dependent choice'' + ``Every real is constructible'' (and therefore
``Well ordering of $\RR$'' and ``Continuum hypothesis'').

\section{The BBC realizability algebra}\label{BBCalg}\noindent
The definition and general properties of realizability algebras are given at the beginning of~\cite{JLK1}. In particular,
closed $\lbd$-terms are interpreted as terms in these algebras.

\smallskip\noindent
The BBC realizability algebra $\gB=(\LLbd,\PPi,\bbot)$ is defined as follows~: 

\smallskip\noindent
$\bullet$~~The set of \emph{processes} $\LLbd\star\PPi$ is $\LLbd\fois\PPi$.

\smallskip\noindent
$\bullet$~~The set of \emph{terms} $\LLbd$ is the smallest set which contains the following
\emph{constants of term}~:

\smallskip
$\BBB,\,\CCC,\,\III,\,\KKK,\,\WWW$ \emph{(Curry's combinators)},  $\ccc$ \emph{(Felleisen-Griffin instruction)},

$\AAA$ \emph{(abort instruction)}, $\pp,\qq_0,\ldots,\qq_N$ \emph{(variables)} where $N$ is a fixed integer~;

\smallskip\noindent
and is such that~:

if $\xi,\eta\in\LLbd$ then $(\xi)\eta\in\LLbd$ \ \emph{(application)}~;

with each sequence $\xi_i(i\in\NN)$ of closed elements of $\LLbd$ (i.e. which contain no variable

$\pp,\qq_0,\ldots,\qq_N$) \ is associated, in a one-to-one (and well founded) way, a constant of term

denoted by $\bigwedge_i\xi_i$.

\smallskip\noindent
Therefore, each term $\xi\in\LLbd$ is a finite sequence of constants of term and parentheses.\\
$\LLbd$ is defined by an induction of length $\aleph_1$ and  is of cardinality $\aleph_1$.

\smallskip\noindent
{\bfseries Notations.}\\
The application $(\ldots((\xi_1)\xi_2)\ldots)\xi_n$ will be often written $(\xi_1)\xi_2\ldots\xi_n$
or even $\xi_1\xi_2\ldots\xi_n$.\\
The finite sequence $\qq_0,\ldots,\qq_N$ will be often written $\vec{\qq}$.

\smallskip\noindent
$\bullet$~~The set of \emph{stacks} $\PPi$ is defined as follows~: a stack $\pi$ is a finite sequence
$t_0\ps\ldots\ps t_{n-1}\ps\pi_0$ with $t_0,\ldots,t_{n-1}\in\LLbd$~; it is terminated by the symbol $\pi_0$ which
represents the \emph{empty stack}.

\smallskip\noindent
For each stack $\pi$, the \emph{continuation} $\kk_\pi$ is a term which is defined by recurrence~:\\
$\kk_{\pi_0}=\AAA$~; \ $\kk_{t\ps\pi}=\ell_t\,\kk_\pi$,
with $\ell_t=((\CCC)(\BBB)\CCC\BBB)t$ or $\lbd k\lbd x(k)(x)t$.\\
Thus, if the stack $\pi$ is $t_0\ps\ldots\ps t_{n-1}\ps\pi_0$, we have~:\\
\centerline{$\kk_\pi=(\ell_{t_0})\ldots(\ell_{t_{n-1}})\AAA$ or $\lbd x(\AAA)(x)t_0\ldots t_{n-1}$.}

\smallskip\noindent
The \emph{integer} $\ul{n}$ is defined as follows~:\\
\centerline{$\ul{0}=(\KKK)\III$ or $\lbd x\lbd y\,y$~; $\ul{n+1}=(\sig)\ul{n}$
with $\sig=(\BBB\WWW)(\CCC)(\BBB)\BBB\BBB$ or $\lbd n\lbd f\lbd x(f)(n)fx$.}

\smallskip\noindent
The relation of \emph{execution} $\succ$ is the least preorder on $\LLbd\star\PPi$ defined by the following rules
(with $\xi,\eta,\zeta\in\LLbd,\pi\in\PPi$ and $n\in\NN$)~:

\smallskip\noindent
1.~~$(\xi)\eta\star\pi\succ\xi\star\eta\ps\pi$~; \ \emph{(push)}\\
2.~~$\BBB\star\xi\ps\eta\ps\zeta\ps\pi\succ\xi\star(\eta)\zeta\ps\pi$~; \ \emph{(apply)}\\
3.~~$\CCC\star\xi\ps\eta\ps\zeta\ps\pi\succ\xi\star\zeta\ps\eta\ps\pi$~; \emph{(switch)}\\
4.~~$\III\star\xi\ps\pi\succ\xi\star\pi$~; \ \emph{(no operation)}\\
5.~~$\KKK\star\xi\ps\eta\ps\pi\succ\xi\star\pi$~; \ \emph{(delete)}\\
6.~~$\WWW\star\xi\ps\eta\ps\pi\succ\xi\star\eta\ps\eta\ps\pi$~; \ \emph{(copy)}\\
7.~~$\ccc\star\xi\ps\pi\succ\xi\star\kk_\pi\ps\pi$~; \ \emph{(save the stack)}\\
8.~~$\AAA\star\xi\ps\pi\succ\xi\star\pi_0$~; \emph{(abort)} or \emph{(delete the stack)}\\
9.~~$\bigwedge_i\xi_i\star\ul{n}\ps\pi\succ\xi_n\star\pi$~; \ \emph{(oracle)}

\smallskip\noindent
When $\xi,\eta\in\LLbd$, we set $\xi\succ\eta$ iff $(\pt\pi\in\PPi)(\xi\star\pi\succ\eta\star\pi)$.

\smallskip\noindent
$\bullet$~~\emph{Proof-like terms}.\\
Let $\PL_0$ be the countable set of terms built with the constants $\BBB,\,\CCC,\,\III,\,\KKK,\,\WWW,\,\ccc$ \
and the application. It is the smallest possible set of \emph{proof-like terms}.\\
We shall also consider the set $\PL$ of \emph{closed terms} (i.e. with no occurrence of $\pp,\vec{\qq}$)
which is of cardinality $\aleph_1$.

\smallskip\noindent
$\bullet$~~\emph{Execution of processes~; definition of $\bbot$}.\\
For every process $\xi\star\pi$, at most one among the rules 1 to 9 applies. By iterating these rules,
we obtain the \emph{reduction} or the \emph{execution} of the process $\xi\star\pi$. This execution stops
if and only if the stack is insufficient (rules 2 to 8) or does not begin with an integer (rule 9) or else
if the process has the form $\pp\star\varpi$ or $\qq_i\star\varpi$.

\smallskip\noindent
Finally, we set  $\bbot=\{\xi\star\pi\in\LLbd\star\PPi\;;\;(\ex\varpi\in\PPi)(\xi\star\pi\succ\pp\star\varpi)\}$.

\begin{lemma}
$\gB$ is a coherent realizability algebra.
\end{lemma}\noindent
\begin{proof}
\emph{$\gB$ is a realizability algebra}~:\\
It remains to check that $\kk_\pi\star\xi\ps\varpi\succ\xi\star\pi$, which is done by recurrence on $\pi$~:\\
if $\pi=\pi_0$, it is rule~8~;\\
if $\pi=t\ps\rho$ we have $\kk_\pi\star\xi\ps\varpi=\kk_{t\ps\rho}\star\xi\ps\varpi=\ell_t\kk_\rho\star\xi\ps\varpi
\succ(\kk_\rho)(\xi)t\star\varpi\succ\kk_\rho\star\xi t\ps\varpi\succ\xi t\star\rho$
(recurrence hypothesis) $\succ\xi\star t\ps\rho$.

\smallskip\noindent
\emph{$\gB$ is coherent}~:\\
If $\theta\in\PL$ then $\theta\star\pi_0\notin\bbot$~; indeed, $\,\pp$ does not appear
during the execution of $\theta\star\pi_0$.
\end{proof}

\subsection*{Models and functionals}
A coherent realizability algebra is useful in order to give \emph{truth values} to formulas of ZF\/. In fact, we use
a theory called \ZFe\ \cite{JLK2} which is a conservative extension of ZF\/.
This theory has an additional strong membership relation symbol $\veps$ \emph{which is not extensional}.

\smallskip\noindent
For each closed formula $F$ of \ZFe, we define \emph{two truth values}, denoted $\|F\|$ and $|F|$, with $\|F\|\subset\PPi$ and 
$|F|\subset\LLbd$, with the relation $\xi\in|F|\Dbfl(\pt\pi\in\|F\|)(\xi\star\pi\in\bbot)$.\\
The relation $\xi\in|F|$ is also written $\xi\forces F$ and reads  \emph{``the term $\xi$ realizes the formula $F$''.}

\smallskip\noindent
All the necessary definitions are given in~\cite{JLK1,JLK2,JLK3}.

\smallskip\noindent
The following lemma~\ref{eta_eq} is a useful property of the BBC realizability algebra $\gB$.

\begin{lemma}\label{eta_eq}
For all formulas $A,B$ of \ \ZFe, and all terms $\,\xi\in\LLbd$, we have~:\\
$\xi\forces A\to B$ \ iff \ $(\pt \eta\in\LLbd)(\eta\forces A\Fl\xi\eta\forces B)$.
\end{lemma}\noindent
Indeed, by the general definition of $\forces$, we have~:\\
\centerline{$(\xi\forces A\to B)\Dbfl(\pt \eta\forces A)(\pt\pi\in\|B\|)(\xi\star\eta\ps\pi\in\bbot)$.}

\noindent
Now, by the above definition of $\bbot$, it is clear that $(\xi\star\eta\ps\pi\in\bbot)\Dbfl(\xi\eta\star\pi\in\bbot)$
from which the result follows.

\qed

\smallskip\noindent
Classical realizability is an extension of forcing. As in forcing, we start with an ordinary model
${\cal M}$ of ZFC (or even ZF + V = L)  which we call the \emph{ground model},
and we build a \emph{realizability model} ${\cal N}$ which satisfies \ZFe\ in the following sense~:\\
${\cal M}$ and ${\cal N}$ have the same domain, but neither the same language, nor the same truth values.
The language of ${\cal N}$ has the additional binary symbol $\veps$ of \emph{strong membership}. The truth
values of ${\cal N}$ are not $0,1$ as for ${\cal M}$, but are taken in ${\cal P}(\PPi)$
endowed with a suitable structure of Bolean algebra \cite{JLK1,JLK3}. We say that \emph{${\cal N}$ satisfies
a formula $F$} iff there is a proof-like term $\theta$ which realizes $F$ or equivalently\/, if the truth
value $\|F\|$ of $F$ is the unit of the Boolean algebra ${\cal P}(\PPi)$.

\smallskip\noindent
A \emph{functional} on the ground model ${\cal M}$ is a formula $F(\vec{x},y)$ of ZF with parameters
in ${\cal M}$, such that ${\cal M}\models\pt\vec{x}\,\ex!y\,F(\vec{x},y)$. Denoting such a functional
by $f$, we write $y=f(\vec{x})$ for $F(\vec{x},y)$.

\smallskip\noindent
Since ${\cal M}$ and ${\cal N}$ have the same domain, all the functionals defined on ${\cal M}$ are also
defined on ${\cal N}$ and \emph{they satisfy the same equations} and even the same \emph{Horn formulas}
i.e. formulas of the form $\pt\vec{x}(f_1(\vec{x})=g_1(\vec{x}),\ldots,f_n(\vec{x})=g_n(\vec{x})\to
f(\vec{x})=g(\vec{x}))$.

\smallskip\noindent
A particularly useful binary functional on ${\cal M}$ (and thus also on ${\cal N}$) is the \emph{application},
denoted by app, which is defined as follows~: \ $\mbox{app}(f,x)=\{y\;;\;(x,y)\in f\}$.\\
We shall often write $f[x]$ for  $\mbox{app}(f,x)$. This allows to consider each set in ${\cal M}$
(and in ${\cal N}$) as a unary functional.

\smallskip\noindent
{\small{\bfseries{}Remark.}
We can define a set $f$ in ${\cal M}$ by giving $f[x]$ for every $x$, provided that there exists
a set $X$ such that $f[x]=\vide$ for all $x\notin X$~: take $f=\bigcup_{x\in X}\{x\}\fois f[x]$.\\
In the ground model ${\cal M}$, every function is defined in this way but in general,
\emph{this is false in ${\cal N}$}.}

\subsection*{Quantifiers restricted to $\NN$}\noindent
In \cite{JLK3}, we defined the quantifier $\pt x\indi$, by setting~:\\
$\|\pt x\indi F[x]\|=\bigcup_{n\in\NN}\|\{\ul{n}\}\to F[n]\}\|=\{\ul{n}\ps\pi\;;\;n\in\NN,\pi\in\|F[n]\|\}$,
so that we have~:\\
$\xi\forces\pt x\indi F[x]\;\Dbfl\;\xi\ul{n}\forces F[n]$ for all $n\in\NN$~;\\
and it is shown that it is a correct definition of the restricted quantifier to $\NN$.\\
Indeed the equivalence \ $\pt x\indi F[x]\dbfl\pt x(\mbox{int}[x]\to F[x])$ is realized by a closed
$\lbd$-term independent of $F$, called a \emph{storage operator}.\\
The formula $\mbox{int}[x]$ is any formula of ZF which says that $x$ is an integer.

\begin{theorem}\label{omega-modele}
If we take $\PL$ for the set of proof-like terms, and if the ground model ${\cal M}$ is transitive and
countable, then there exists a countable realizability model ${\cal N}$ which has only standard integers,
i.e. which is an $\omega$-model.
\end{theorem}\noindent
Let ${\cal T}$ be the theory formed with closed formulas, with parameters in ${\cal M}$, which are
realized by a proof-like term. ${\cal T}$ is $\omega$-complete~: indeed, if $\theta_n\in\PL$ and
$\theta_n\force F[n]$ for $n\in\NN$, let us set $c=\bigwedge_i\theta_i$. Then $c\ul{n}\force F[n]$
for all $n\in\NN$ and therefore $c\force\pt n\indi F[n]$, i.e. $\pt n\indi F[n]\in{\cal T}$.
It follows that ${\cal T}$ has a countable $\omega$-model.

\qed

\begin{proposition}\ \\
Let $f:\NN\to2$ and $\theta\in\PL$, $\theta\forces\ex n\indi(f(n)=1)$.
Then $\;\theta\star\pp\ps\pi_0\succ\pp\star\ul{n}\ps\varpi$ with $f(n)=1$.
\end{proposition}\noindent
There exists $\tau\in\LLbd$ such that $\tau\ul{n}\succ\pp$ if $f(n)=1$ and $\tau\ul{n}\succ\qq_0$ if $f(n)=0$~:
set $\tau=\lbd x(\bigwedge_i\xi_i)x\,\pp\qq_0$ with $\xi_n=\KKK$ if $f(n)=1$ and $\xi_n=\KKK\III$ if $f(n)=0$.\\
Then we have $\tau\forces\pt n\indi(f(n)\ne1)$ and therefore $\theta\tau\forces\bot$. We necessarily have~:\\
$\theta\star\tau\ps\pi_0\succ\tau\star\ul{n}\ps\pi$ for some $n$~; furthermore, we have $\tau\ul{n}\succ\pp$,
otherwise we should have $\tau\ul{n}\succ\qq_0$, and thus $\;\theta\star\tau\ps\pi_0\notin\bbot$.
Therefore $f(n)=1$.

\qed

\smallskip\noindent
{\small{\bfseries{}Remark.}
This shows that, from any proof-like term which realizes a given $\Sigma^0_1$ arithmetical formula, we obtain
a program which computes an integer satisfying this formula. Such a realizer is given by any proof of this formula by means of axioms which have themselves such realizers.\\
The theory of classical realizability gives realizers for the axioms of ZF\/. We show below that the bar recursion operator
realizes the axiom of dependent choice. Finally, in section~\ref{WOR}, we get (rather complicated) proof-like realizers
for the axioms ``$\RR$ is well ordered'' and ``Continuum hypothesis''.}

\subsection*{Execution of processes}

{\bfseries Notation.} If $\pi=t_0\ps\ldots\ps t_{n-1}\ps\pi_0$, we shall write
$\pi\ps t$ for $t_0\ps\ldots\ps t_{n-1}\ps t\ps\pi_0$.\\
Thus, we obtain $\kk_{\pi\ps t}$ by replacing, in $\kk_\pi$, the last occurrence of $\AAA$ by $\ell_t\AAA$.

\begin{lemma}\label{pi-ps-t}
If $\,\xi\star\pi\in\bbot$, then $\,\xi'\star\pi'\in\bbot$ and $\,\xi'\star\pi'\ps t\in\bbot$, where $\,\xi'\star\pi'$ is obtained by replacing, in $\,\xi\star\pi$, some occurrences of $\,\AAA$ by $(\ell_u)\AAA=\kk_{u\ps\pi_0}$ and some occurrences of the variabless
$\,\qq_0,\ldots,\qq_N$ by $\,t_0,\ldots,t_N$~; $t_0,\ldots,t_N,t,u$ are arbitrary terms.
\end{lemma}\noindent
{\small{\bfseries{}Remark.}
In particular, it follows that $\,\xi\star\pi_0\in\bbot\Fl\xi\forces\bot$.}

\smallskip\noindent
Proof by recurrence on the length of the execution of $\xi\star\pi\in\bbot$ by means of rules 1 to~9.
We consider the last used rule. There are two non trivial cases~:

\smallskip\noindent
$\bullet$~~Rule~7 (execution of $\ccc$)~; we must show
$\ccc\star\xi'\ps\pi'\ps t\in\bbot$.\\
We apply the recurrence hypothesis to $\,\xi\star\kk_\pi\ps\pi$, in which we replace~:\\
- $\pi_0$ by $t\ps\pi_0$ (thus $\pi$ becomes $\pi\ps t$)~;\\
- the last occurrence of $\AAA$ in $\kk_\pi=(\ell_{t_0})\ldots(\ell_{t_{n-1}})\AAA$ by $(\ell_t)\AAA$
(thus $\kk_\pi$ becomes $\kk_{\pi\ps t}$).\\
Then, we make the substitutions in $\xi,\pi$, which gives $\xi'\star\kk_{\pi'\ps t}\ps\pi'\ps t$.

\smallskip\noindent
$\bullet$~~Rule~8 (execution of $\AAA$)~; we must show $(\ell_u)\AAA\star\xi'\ps\pi'\ps t\in\bbot$.\\
We apply the recurrence hypothesis to $\,\xi\star\pi_0$, which gives $\xi'\star u\ps\pi_0\in\bbot$, thus
$\xi' u\star\pi_0\in\bbot$ and therefore $\AAA\star\xi'u\ps\pi'\ps t\in\bbot$ (rule~8)~; finally, we obtain
$(\ell_u)\AAA\star\xi'\ps\pi'\ps t\in\bbot$.

\qed

\smallskip\noindent
In each process $\,\xi\star\pi\in\bbot$, we define an occurrence of $\pp$, which is called \emph{efficient},
by recurrence on the length of its reduction. If $\xi=\pp$, it is this very occurrence. Otherwise, we
consider the last rule used in the reduction, and the definition is clear~; for example, if it is rule~7,
and if the efficient occurrence in $\xi\star\kk_\pi\ps\pi$ is in $\kk_\pi$ or in $\pi$, then we take
the corresponding occurrence in $\ccc\star\xi\ps\pi$.

\begin{lemma}\label{efficace}
If $\;\xi\star\pi\in\bbot$, then~:
\begin{itemize}
\item $\xi'\star\pi'\in\bbot$, where $\xi'\star\pi'$ is obtained by substituting arbitrary terms for
the non efficient occurrences of $\,\pp$.
\item $\xi'\star\pi'\notin\bbot$ and indeed $\xi'\star\pi'\succ\qq_0\star\varpi$, where $\xi'\star\pi'$
is obtained by substituting $\;\qq_0$ for the efficient occurrence of $\,\pp$, and arbitrary terms for
the non efficient occurrences of $\,\pp$.
\end{itemize}
\end{lemma}\noindent
The proof is immediate, by recurrence on the length of the reduction of
$\xi\star\pi$ by means of rules 1 to 9~: consider the last used rule.

\qed

\begin{corollary}\label{gl2nt}\ \\
If $\xi\forces\top,\bot\to\bot$ and $\xi\forces\bot,\top\to\bot$, then $\xi\forces\top,\top\to\bot$,
and thus~:\\
$\lbd x(x)\III\,\III\,\forces\neg\pt x^{\gl2}(x\ne0,x\ne1\to\bot)$ and\\
$\WWW\,\forces\pt x^{\gl2}(\pt y^{\gl2}(y\ne0,y\ne x\to y\not\le x),x\ne0\to\bot)$.
\end{corollary}\noindent
{\small{\bfseries{}Remark.} These two formulas express respectively that the Boolean algebra $\gl2$ is
non trivial and that it is atomless.}

\smallskip\noindent
We apply lemma~\ref{efficace} to $\xi\star\pp\ps\pp\ps\pi_0$. We have $\xi\star\qq_0\ps\pp\ps\pi_0\in\bbot$
and $\xi\star\pp\ps\qq_0\ps\pi_0\in\bbot$, which shows that the efficient occurrence of $\pp$ is in $\xi$.
Therefore $\xi\star t\ps u\ps\pi_0\in\bbot$ for every $t,u\in\LLbd$, again by lemma~\ref{efficace}.\\
The last two assertions follow from the fact that~:\\
$\|\pt x^{\gl2}(x\ne0,x\ne1\to\bot)\|=\|\top,\bot\to\bot\|\cup\|\bot,\top\to\bot\|$ and therefore~:\\
$|\pt x^{\gl2}(x\ne0,x\ne1\to\bot)|=|\top,\top\to\bot|$.

\qed

\begin{theorem}\label{den_cont}
For every sequence $\xi_i\in\LLbd$ $(i\in\NN)$, there exists $\;\phi\in\LLbd$ such that~:\\
$\bullet$~~$\phi\ul{i}\succ\xi_i$ for every $i\in\NN$~;\\
$\bullet$~~for every $\;U\in\LLbd$ such that $\;U\phi\forces\bot$, there exists $k\in\NN$ such that
$\;U\psi\forces\bot$ for every $\;\psi\in\LLbd$ such that \ $\psi\ul{i}\succ\xi_i$ for every $i<k$.
\end{theorem}\noindent
{\small{\bfseries{}Remark.} Theorem~\ref{den_cont} will be used in order to show properties of the
bar recursion operator.  In fact, the following weaker formulation is sufficient~:

\noindent
For every sequence $\xi_i\in\LLbd$ $(i\in\NN)$ and every $U\in\LLbd$ such that~:\\
\centerline{$(\pt k\in\NN)(\ex\psi\in\LLbd)\{U\psi\nforces\bot,(\pt i<k)(\psi\ul{i}\succ\xi_i)\}$}

\noindent
there exists $\phi\in\LLbd$ such that $U\phi\nforces\bot$ and $(\pt i\in\NN)(\phi\ul{i}\succ\xi_i)$.\\
In the particular case of forcing, this is exactly the \emph{decreasing chain condition}~:
every decreasing sequence of (non false) conditions has a lower bound (which is non false).}

\smallskip\noindent
We set $\eta_i=\lbd\pp\lbd\vec{\qq}\,\xi_i$~; thus, we have $\eta_i\in\PL$ and $\eta_i\pp\vec{\qq}\succ\xi_i$.\\
Let $\eta=\bigwedge_i\eta_i$ and $\phi=\lbd x(\eta)x\,\pp\vec{\qq}$.
Thus, we have $\eta\in\PL$ and $\phi\ul{i}\succ\xi_i$.\\
We may assume that $\eta$ does not appear in $U$.\\
We have $U\phi\forces\bot\Dbfl U\star\phi\ps\pi_0\in\bbot$ (lemma~\ref{pi-ps-t}).
During the execution of the process $U\star\phi\ps\pi_0$, the constant $\eta$ arrives in head position a finite
number of times, always through~$\phi$ (since it is deleted each time it arrives in head position),
therefore as follows~:\\
\centerline{$\eta\star\ul{i}\ps\pp\ps\vec{\qq}\ps\pi\succ\xi_i\star\pi$.}

\noindent
Let $k$ be an integer, greater than all the arguments of $\eta$ during this execution and let $\;\psi\in\LLbd$
be such that $\,\psi\ul{i}\succ\xi_i$ for all $i<k$. Let us set $\tau=\lbd x\lbd\pp\lbd\vec{\qq}\;\psi x$~;
thus, we have $\tau\ul{i}\,\pp\vec{\qq}\succ\,\psi\ul{i}\succ\xi_i$ for $i<k$.
In the process $U\star\phi\ps\pi_0$, let us replace the constant $\eta$ by the term $\tau$~; we obtain
$U\star\psi\ps\pi_0$. The execution is the same, and therefore $U\star\psi\ps\pi_0\in\bbot$ and $U\psi\forces\bot$.

\qed

\subsection*{The bar recursion operator}\noindent
We define below two \emph{proof-like terms} $\chi$ and $\Psi$ (which are, in fact, closed $\lbd$-terms).\\
In these definitions, the variables $i,k$ represent (intuitively) integers and the variable~$f$
represents a function of domain $\NN$, with arbitrary values in $\LLbd$.

\smallskip\noindent
$\bullet$~~We want a $\lbd$-term $\chi$ such that~:

\smallskip
\centerline{$\chi\ul{k}fz\ul{i}\succ f\ul{i}$ \ if \ $i<k$~; $\chi\ul{k}fz\ul{i}\succ z$ \ if \ $i\ge k$.}

\smallskip\noindent
Therefore, we set~:\\
\centerline{$\chi=\lbd k\lbd f\lbd z\lbd i((i\ppt k)(f)i)z$}

\noindent
where the boolean $(i\ppt k)$ is defined by~:

\noindent\centerline{
$(i\ppt k)=((kA)\lbd d\,\mathbf{0})(iA)\lbd d\,\mathbf{1}$}

\noindent
with $\mathbf{0}=\lbd x\lbd y\,y$ or $\KKK\,\III$, $\mathbf{1}=\lbd x\lbd y\,x$ or $\KKK$ and
$A=\lbd x\lbd y\,yx$ or $\CCC\,\III$.

\smallskip\noindent
The term $\chi\ul{k}f$ is a representation, in $\lbd$-calculus, of the finite sequence
$(f\ul{0},f\ul{1},\ldots,f\ul{k-1})$.

\smallskip\noindent
$\bullet$~~We want a $\lbd$-term $\Psi$ such that~:

\smallskip
\centerline{$\Psi gu\ul{k}f\succ(u)(\chi\ul{k}f)(g)\lbd z(\Psi gu\ul{k}^+)(\chi)\ul{k}fz$}

\smallskip\noindent
where $k^+=((\BBB\WWW)(\CCC)(\BBB)\BBB\BBB)k$ or $\lbd f\lbd x(f)(k)fx$ is the successor of the integer $k$.\\
Thus, we set~:

\smallskip\noindent
\centerline{$\Psi=\lbd g\lbd u(\Y)\lbd h\lbd k\lbd f(u)(\chi kf)(g)\lbd z(hk^+)(\chi)kfz$.}

\smallskip\noindent
where $\Y$ is the Turing fix point operator~:\\
\centerline{$\Y=XX$ \ with $X=\lbd x\lbd f(f)(x)xf=(\WWW)(\BBB)(\BBB\WWW)(\CCC)\BBB$.}

\smallskip\noindent
The term $\Psi$ will be called the \emph{bar recursion operator}.

\section{Realizing countable choice}\label{chden}\noindent
The \emph{axiom of countable choice} is the following formula~:

\smallskip\noindent
(CC)\hspace{7em} $\pt n\ex x\, F[n,x]\to\ex f\pt n\indi F[n,f[n]]$

\smallskip\noindent
where $F[n,x]$ is an arbitrary formula
of \ZFe(see~\cite{JLK2}), with parameters and two free variables.
The notation $f[n]$ stands for $\mbox{app}(f,n)$ (the functional app has been defined above).

\smallskip\noindent 
{\small{\bfseries{}Remark.}
This is a strong form of countable choice which shows that, in the realizability model~${\cal N}$, every
countable sequence has the form $n\mapsto f[n]$ for some $f$. This will be used in section~\ref{WOR}.}

\begin{theorem}\label{CC}
$\lbd g\lbd u(\Psi)gu\,\ul{0}\,\ul{0}\,\forces\mbox{\rm CC}$.
\end{theorem}\noindent
The axiom of countable choice is therefore realized in the model of ZF associated with the BBC realizability algebra
(in fact, it is sufficient that the realizability algebra satisfies the property formulated in the remark following
theorem~\ref{den_cont}).

\smallskip\noindent
We write the axiom of countable choice as follows~:

\smallskip\noindent
(CC)\hspace{7em} $\pt n\neg\pt x\neg F[n,x],
\pt f\neg\pt n\indi F[n,f[n]]\to\bot$

\smallskip\noindent
Let $G,U\in\LLbd$ be such that \ $G\forces\pt n\neg\pt x\neg F[n,x]$ \ and \
$U\forces\pt f\neg\pt n\indi F[n,f[n]]$.\\
We set $H=\Psi GU$ and we have to show that $H\ul{0}\,\ul{0}\forces\bot$. In fact, we shall show that \
$H\ul{0}\xi\forces\bot$ for every $\xi\in\LLbd$.

\begin{lemma}\label{rec-Psi2}
Let $k\in\NN$ and \ $\phi\in\LLbd$ be such that \ $(\pt i<k)\ex a_i(\phi \ul{i}\forces F[i,a_i])$.\\
If $H\ul{k}\,\phi\nforces\bot$, then there exist a set $a_k$ and a term $\;\zeta_{k,\;\phi}\in\LLbd$ such that~:\\
\centerline{$\zeta_{k,\;\phi}\forces F[k,a_k]$ \ and \ $(H\ul{k}^+)(\chi)\ul{k}\,\phi\,\zeta_{k,\;\phi}\nforces\bot$.}
\end{lemma}\noindent
Define $\eta_{k,\;\phi}=\lbd z(H\ul{k}^+)(\chi)\ul{k}\,\phi z$, so that
$H\ul{k}\,\phi\succ(U)(\chi\ul{k}\,\phi)(G)\eta_{k,\;\phi}$.\\
If \ $\eta_{k,\;\phi}\forces\pt x\neg F[k,x]$ \ then, by hypothesis on $G$, we have $G\eta_{k,\;\phi}\forces\bot$.
Let us check that~:\\
\centerline{$(\chi\ul{k}\,\phi)(G)\eta_{k,\;\phi}\forces\pt n\indi F[n,f_k[n]]$}

\noindent
where $f_k$ is defined by~: \ $f_k[i]=a_i$ if $i<k$ (i.e. $i\in k$)~; $f_k[i]=\vide$ if $i\notin k$.\\
Indeed, if we set \ $\phi'=(\chi\ul{k}\phi)(G)\eta_{k,\;\phi}$, we have~:\\
$\phi'\ul{i}\succ\phi\ul{i}\forces F[i,a_i]$ for $i<k$ and $\phi'\ul{i}\succ(G)\eta_{k,\;\phi}\forces\bot$
for $i\ge k$, and therefore $\phi'\ul{i}\forces F[i,\vide]$.\\
By hypothesis on $U$, it follows that $(U)(\chi\ul{k}\,\phi)(G)\eta_{k,\;\phi}\forces\bot$, in other words $H\ul{k}\,\phi\forces\bot$.\\
Thus, we have shown that, if $H\ul{k}\,\phi\nforces\bot$, then $\eta_{k,\;\phi}\nforces\pt x\neg F[k,x]$,
which gives immediately the desired result.

\qed

\smallskip\noindent
Let $\phi_0\in\LLbd$ be such that $H\ul{0}\phi_0\nforces\bot$. By means of lemma~\ref{rec-Psi2},
we define $\phi_{k+1}\in\LLbd$ and $a_k$ recursively on $k$, by setting \
$\phi_{k+1}=\chi\ul{k}\,\phi_k\,\zeta_{k,\;\phi_k}$.\\
By definition of $\chi$, we have $\phi_{k+1}\ul{i}\succ\zeta_{k,\phi_k}$ for $i\ge k$.\\
Then, we show easily, by recurrence on $k$~:

\smallskip\centerline{
$\phi_{k+1}\ul{i}\succ\phi_{i+1}\ul{i}\succ\zeta_{i,\phi_i}\forces F[i,a_i]$ \ for $i\le k$~; \ $H\ul{k}\phi_k\nforces\bot$.}

\smallskip\noindent
Therefore, we can define~:\\
a function $f$ of domain $\NN$ such that $f[i]=a_i$ for every $i\in\NN$~;\\
and, by theorem~\ref{den_cont}, a term $\phi\in\LLbd$ such that $\phi\ul{k}\succ\zeta_{k,\;\phi_k}$
for all $k\in\NN$.\\
Therefore, we have $\phi\ul{i}\forces F[i,f[i]]$ for every $i\in\NN$, that is to say \
$\phi\forces\pt n\indi F[n,f[n]]$.\\
By hypothesis on $U$, it follows that \ $U\phi\forces\bot$.
Therefore, by theorem~\ref{den_cont}, applied to the sequence $\xi_i=\zeta_{i,\;\phi_i}$, there exists an
integer~$k$ such that $U\psi\forces\bot$, for every
term $\psi\in\LLbd$ such that $\psi\ul{i}\succ\zeta_{i,\;\phi_i}$ for $i<k$.

\smallskip\noindent
Thus, in particular, we have $(U)(\chi\ul{k}\,\phi_k)\xi\forces\bot$ for every $\xi\in\LLbd$.

\smallskip\noindent
Now, by definition of $H$, we have \ $H\ul{k}\,\phi_k\succ(U)(\chi\ul{k}\,\phi_k)\xi$ with
$\xi=(G)\lbd z(H\ul{k}^+)(\chi)\ul{k}\,\phi_kz$, and therefore \ $H\ul{k}\,\phi_k\forces\bot$,
that is a contradiction.

\smallskip\noindent
Thus, we have shown that $H\ul{0}\phi_0\forces\bot$ for every $\phi_0\in\LLbd$.

\qed

\section{Realizing dependent choice}\noindent\label{chdep}
The \emph{axiom of dependent choice} is the following formula~:\\
(DC)\hspace{8em} $\pt x\ex y\,F[x,y]\to\ex f\pt n\indi F[f[n],f[n+1]]$\\
where $F[x,y]$ is an arbitrary formula of \ZFe, with parameters and two free variables.\\
The notation $f[n]$ stands for $\mbox{app}(f,n)$ as defined above.

\begin{theorem}\label{DC}
$\lbd g\lbd u(\Psi)gu\,\ul{0}\,\ul{0}\forces\mbox{\rm DC}$. 
\end{theorem}\noindent
The axiom of dependent choice is therefore realized in the model of ZF associated with the BBC realizability algebra
(or, more generally, with any realizability algebra satisfying the property formulated in the remark after
theorem~\ref{den_cont}).

\smallskip\noindent
The proof of theorem~\ref{DC} is almost the same as theorem~\ref{CC}.\\
We write the \emph{axiom of dependent choice} as follows~:\\
(DC)\hspace{8em} $\pt x\,\neg\pt y\,\neg F[x,y],\,\pt f\neg\pt n\indi F[f[n],f[n+1]]\to\bot$.

\smallskip\noindent
Let $G,U\in\LLbd$ be such that \ $G\forces\pt x\,\neg\pt y\,\neg F[x,y]$ and
$U\forces\pt f\neg\pt n\indi F[f[n],f[n+1]]$.\\
We set $H=\Psi GU$ and we have to show that $H\ul{0}\,\ul{0}\forces\bot$. In fact, we shall show that \
$H\ul{0}\xi\forces\bot$ for every $\xi\in\LLbd$.

\begin{lemma}\label{rec-Psi3}\ \\
Let $a_0,\ldots,a_k$ be a finite sequence in ${\cal M}$ and $\phi\in\LLbd$ be such that \
$(\pt i<k)(\phi \ul{i}\forces F[a_i,a_{i+1}])$.\\
If $H\ul{k}\,\phi\nforces\bot$, then there exist $\zeta\in\LLbd$ and $a_{k+1}$ in ${\cal M}$ such that~:\\
\centerline{$\zeta\forces F[a_k,a_{k+1}]$ and $(H\ul{k}^+)(\chi)\ul{k}\,\phi\,\zeta\nforces\bot$.}
\end{lemma}\noindent
Define $\eta_{k,\;\phi}=\lbd z(H\ul{k}^+)(\chi)\ul{k}\,\phi z$, so that
$H\ul{k}\,\phi\succ(U)(\chi\ul{k}\,\phi)(G)\eta_{k,\;\phi}$.\\
If \ $\eta_{k,\;\phi}\forces\pt y\,\neg F[a_k,y]$ \ then, by hypothesis on $G$, we have
$G\eta_{k,\;\phi}\forces\bot$. We check that~:\\
\centerline{$(\chi\ul{k}\,\phi)(G)\eta_{k,\;\phi}\forces\pt n\indi F[f_k[n],f_k[n+1]]$}

\noindent
where $f_k$ is defined by $f_k[i]=a_i$ for $i\le k$ (i.e. $i\in k+1$)~; $f_k[i]=\vide$ for $i\notin k+1$.\\
Indeed, if we set \ $\phi'=(\chi k\phi)(G)\eta_{k,\;\phi}$, we have~:\\
$\phi'\ul{i}\succ\phi\ul{i}\forces F[a_i,a_{i+1}]$ for $i<k$ and
$\phi'\ul{i}\succ(G)\eta_{k,\;\phi}\forces\bot$ for $i\ge k$.\\
Therefore, we have $\phi'\ul{i}\forces F[f_k[i],f_k[i+1]]$ for every $i\in\NN$.

\smallskip\noindent
By hypothesis on $U$, it follows that $(U)(\chi\ul{k}\,\phi)(G)\eta_{k,\;\phi}\forces\bot$, that is
$H\ul{k}\,\phi\forces\bot$.

\smallskip\noindent
Thus, we have shown that, if $H\ul{k}\,\phi\nforces\bot$, then $\eta_{k,\;\phi}\nforces\pt y\,\neg F[a_k,y]$,
which gives immediately the desired result.

\qed

\smallskip\noindent
Let $\phi_0\in\LLbd$ be such that $H\ul{0}\phi_0\nforces\bot$ and let $a_0=\vide$. Using lemma~\ref{rec-Psi3},
we define $\phi_{k+1}\in\LLbd$ and $a_{k+1}$ in ${\cal M}$ recursively on $k$, by setting \
$\phi_{k+1}=\chi\ul{k}\,\phi_k\,\zeta_{k,\,\phi_k}$, where $\zeta_{k,\,\phi_k}$ is given by lemma~\ref{rec-Psi3},
where we set $\phi=\phi_k$.  By definition of $\chi$, we have $\phi_{k+1}\ul{i}\succ\zeta_{k,\phi_k}$ for $i\ge k$.\\
Then, we show easily, by recurrence on $k$~:

\smallskip\centerline{
$\phi_{k+1}\ul{i}\succ\phi_{i+1}\ul{i}\succ\zeta_{i,\phi_i}\forces F[a_i,a_{i+1}]$ \ for $i\le k$~; \
$H\ul{k}\phi_k\nforces\bot$.}

\smallskip\noindent
Therefore, we can define~:\\
a function $f$ of domain $\NN$ such that $f[i]=a_i$ for every $i\in\NN$~;\\
and, by means of theorem~\ref{den_cont}, a term $\phi\in\LLbd$ such that
$\phi\ul{k}\succ\zeta_{k,\,\phi_k}$ for every $k\in\NN$.\\
Thus, we have $\phi\ul{i}\forces F[f[i],f[i+1]]$ for every $i\in\NN$, that is to say \
$\phi\forces\pt n\indi F[f[n],f[n+1]]$.\\
By hypothesis on $U$, it follows that \ $U\phi\forces\bot$.
Therefore, by theorem~\ref{den_cont}, applied to the sequence $\xi_i=\zeta_{i,\,\phi_i}$,
there exists an integer~$k$ such that $U\psi\forces\bot$, for every
term $\psi\in\LLbd$ such that $\psi\ul{i}\succ\zeta_{i,\,\phi_i}$ for $i<k$.

\smallskip\noindent
Thus, in particular, we have $(U)(\chi\ul{k}\,\phi_k)\xi\forces\bot$ for every $\xi\in\LLbd$.

\smallskip\noindent
But, by definition of $H$, we have \ $H\ul{k}\,\phi_k\succ(U)(\chi\ul{k}\,\phi_k)\xi$ with
$\xi=(G)\lbd z(H\ul{k}^+)(\chi)\ul{k}\,\phi_kz$, and therefore \ $H\ul{k}\,\phi_k\forces\bot$,
that is a contradiction.

\smallskip\noindent
Thus, we have shown that $H\ul{0}\phi_0\forces\bot$ for every $\phi_0\in\LLbd$.

\qed

\section{A well ordering on $\mathbb{R}$}\label{WOR}\noindent
In this section, we use the notations and the results of \cite{JLK3} and \cite{JLK4}.\\
If $F$ is a closed formula of \ZFe, the notation $\forces F$ \ means that there exists a proof-like
term \ $\theta\in\PL_0$ (i.e. a closed $\lbd_c$-term) such that $\theta\forces F$.

\smallskip\noindent
In section~\ref{chden}, we have realized the axiom of countable choice (CC).
We replace $F[n,x]$ with $\mbox{int}(n)\to F[n,x]$ and we add a parameter $\phi$~; we obtain~:

\smallskip
\centerline{$\forces\pt \phi\left(\pt n\indi\ex x\,F[n,x,\phi]\to
\ex f\pt n\indi F[n,f[n],\phi]\right)$}

\smallskip\noindent
for every formula $F[n,x,\phi]$ of \ZFe.

\smallskip\noindent
In particular, taking $\phi\eps2^\NN$ and $F[n,x,\phi]\equiv(x=\phi(n))\land(x=0\lor x=1)$\\
(i.e. $(n,x)\eps\phi\land(x=0\lor x=1)$), we find~:

\smallskip
\centerline{$\forces(\pt \phi\eps2^\NN)\ex f\pt n\indi\left((f[n]=\phi(n))\land(f[n]=0\,\lor f[n]=1)\right)$.}

\smallskip\noindent
For any set $f$ in the ground model ${\cal M}$, let $g=\{x\;;\;f[x]=1\}$.\\
We have trivially \ $\III\,\forces\lbr n\in g\rbr=\lbr f[n]=1\rbr$.%
\footnote{The notations $\gl2$ and $\lbr F\rbr$ where $F$ is a closed formula of ZF\/,
with parameters in the realizability model ${\cal N}$, are defined in \cite{JLK3,JLK4}.
$\gl2$ is called the \emph{characteristic Boolean algebra of ${\cal N}$}. We have $\lbr F\rbr\eps\gl2$.}\\
It follows that~: $\forces\pt f\ex g\pt n\left((f[n]=0\,\lor f[n]=1)\to f[n]=\lbr n\in g\rbr\right)$.

\smallskip\noindent
We have shown that~: $\forces(\pt \phi\eps2^\NN)\ex g\pt n\indi(\phi(n)=\lbr n\in g\rbr)$.

\smallskip\noindent
Now, in \cite{JLK4}, we have built an ultrafilter ${\cal D}:\gl2\to2$ on the Boolean algebra $\gl2$,
with the following property~: the model ${\cal N}$, equipped with the binary relations
${\cal D}(\lbr x\in y\rbr)$, ${\cal D}(\lbr x=y\rbr)$, is a model of ZF\/, denoted ${\cal M}_{\cal D}$,
which is an elementary extension of the ground model ${\cal M}$. Moreover, ${\cal M}_{\cal D}$ is isomorphic
to a transitive submodel of ${\cal N}$ (considered as a model of ZF), which contains every ordinal of~${\cal N}$.\\
${\cal M}_{\cal D}$ satisfies the axiom of choice, because we suppose that ${\cal M}\models \mbox{ZFC}$.\\
If we suppose that ${\cal M}\models\mbox{ V = L}$, then ${\cal M}_{\cal D}$ is isomorphic to the class $L^{\cal N}$
of constructible sets of~${\cal N}$.

\smallskip\noindent
For every $\phi:\NN\to2$, we have obviously ${\cal D}(\phi(n))=\phi(n)$. It follows that~:

\smallskip
\centerline{$\forces(\pt \phi\eps2^\NN)\ex g\pt n\indi\left(\phi(n)={\cal D}\lbr n\in g\rbr\right)$.}

\smallskip\noindent
This shows that the subset of $\NN$ defined by $\phi$ is in the model ${\cal M}_{\cal D}$~:
indeed, it is the element $g$ of this model.\\
We have just shown that \emph{${\cal N}$ and ${\cal M}_{\cal D}$ have the same reals}.\\
Therefore, $\mathbb{R}$ is well ordered in~${\cal N}$, and we have~: $\forces(\mathbb{R}$ \emph{is well ordered}).

\smallskip\noindent
Moreover, if the ground model ${\cal M}$ satisfies V = L, we have~: $\forces$ (\emph{every real is
constructible}). Therefore, the continuum hypothesis is realized.

\smallskip\noindent
Since the models ${\cal N}$ and ${\cal M}_{\cal D}$ have the same reals, every formula of analysis
(closed formula with quantifiers restricted to $\NN$ or $\mathbb{R}$) has the same truth value in
${\cal M}_{\cal D},\,{\cal M}$ or ${\cal N}$.\\
It follows that~:

\smallskip\noindent
\emph{For every formula $F$ of analysis, we have \ ${\cal M}\models F$ \ if and only if \ $\forces F$.}\\
In particular, we have $\forces F$ or $\forces\neg F$.

\noindent
References~\cite{JLK1,JLK2,JLK3,JLK4} are available at
{\tt www.irif.univ-paris-diderot.fr/$\!\sim$krivine/ }

\end{document}